\documentclass{emulateapj}

\newcommand{\etal}{{\it et.al\,}}
\newcommand{\eg}{{\it eg.,\ }}
\newcommand{\ie}{{\it ie.,\ }}

\shorttitle{Gravitational waves from supernovae}
\shortauthors{M\"uller et.al}

\begin{document}

\title{Towards gravitational wave signals from realistic core collapse
       supernova models}

\author{Ewald M\"uller, Markus Rampp, Robert Buras, H.-Thomas Janka}
\affil{Max-Planck-Institut f\"ur Astrophysik,
       Karl-Schwarzschild-Str.\ 1, D-85740 Garching, Germany}
\and
\author{David H.\,Shoemaker}
\affil{LIGO Project, 175 St.\,Albany Street, Massachusetts 
       Institute of Technology, Cambridge, MA 02139, USA\\}

\begin{abstract}
We have computed the gravitational wave signal from supernova core
collapse using the presently most realistic input physics
available. We start from state-of-the-art progenitor models of
rotating and non-rotating massive stars, and simulate the dynamics of
their core collapse by integrating the equations of axisymmetric
hydrodynamics together with the Boltzmann equation for the neutrino
transport including an elaborate description of neutrino interactions,
and a realistic equation of state.  Using the Einstein quadrupole
formula we compute the quadrupole wave amplitudes, the Fourier wave
spectra, the amount of energy radiated in form of gravitational waves,
and the signal-to-noise ratios for the LIGO\,I and the tuned Advanced
LIGO (``LIGO\,II'') interferometers resulting both from non-radial
mass motion and anisotropic neutrino emission.  The simulations
demonstrate that the dominant contribution to the gravitational-wave
signal is produced by neutrino-driven convection behind the supernova
shock. For stellar cores rotating at the extreme of current stellar
evolution predictions, the core-bounce signal is detectable ($S/N
\ga 7$) with LIGO\,II for a supernova up to a distance of $\sim
5\,$kpc, whereas the signal from post-shock convection is observable
($S/N \ga 7$) with LIGO\,II up to a distance of $\sim 100\,$kpc, and
with LIGO\,I to a distance of $\sim 5\,$kpc.  If the core is 
non-rotating its gravitational wave emission can be measured with LIGO\,II
up to a distance of $\sim 15\,$kpc ($S/N \ga 8$), while the signal
from the Ledoux convection in the deleptonizing, nascent neutron star
can be detected up to a distance of $\sim 10\,$kpc ($S/N \ga 8$).
Both kinds of signals are generically produced by convection in any
core collapse supernova.

\end{abstract}


\maketitle

\section{Introduction}
For more than two decades astrophysicists struggle to compute the
gravitational wave signal produced by core collapse supernovae (for a
review, see \eg M\"uller 1997).  If the core collapse and/or the
resulting supernova explosion involves time-dependent asphericities
such that the third time-derivative of the quadrupole moment of the
mass-energy distribution is nonzero, part of the gravitational binding
energy liberated in the event will be emitted in the form of
gravitational waves.  Such non-sphericities can be caused on large
scales by the effects of rotation \citep{mueller82, finn90,
moenchmeyer91, yamada95, zwerger97, rampp98, dimmelmeier02, fryer02,
fryer03, imamura03, ott03, shibata03} and low-mode convection
\citep{herant95, scheck03}, on small scales by flow fluctuations due
to high-mode convection \citep{muejan97, fryer03}, and by anisotropic
neutrino emission \citep{epstein78, burrows96, muejan97}.

Theoretical predictions of the gravitational wave signal from core
collapse supernovae have been, and still are, hampered by the complex
nature of the supernova explosion physics which is not yet fully
understood \citep{buras03}.  Reliable simulations require realistic
(rotating) pre-collapse stellar models, the incorporation of a
realistic equation of state (EoS), a detailed modeling of weak
interaction processes, Boltzmann neutrino transport, multi-dimensional
hydrodynamics, and relativistic gravity.  However, all past studies
aimed at studying the gravitational wave signature of core collapse
supernovae have considered greatly simplified parameterized models
involving one, several or all of the following approximations: A
polytropic equation of state, a simplified description of weak
interactions and neutrino transport or none at all, parameterized
pre-collapse stellar models, and Newtonian gravity \citep{mueller82,
finn90, moenchmeyer91, yamada95, zwerger97, rampp98, dimmelmeier02,
fryer02, fryer03, imamura03, kotake03, ott03, shibata03}.  In
addition, most of these simulations focussed on the effects of rapid
rotation, and on the bounce signal covering only the evolution up to a
few ten milliseconds after core bounce.

From observations as well as theoretical modeling it is now commonly
accepted that core collapse supernovae do generically involve
asphericities besides those expected in the case of a progenitor with
a sufficiently rapidly rotating core (for a review see, \eg
\citet{mueller98}). These asphericities are important or may be even
essential for supernova dynamics \citep{herant94, burrows95, janmue96,
buras03}. Recent state-of-the-art axisymmetric (2D) simulations using
a Boltzmann solver for the $\nu$ transport \citep{ramjan02, buras03}
and simplified (as far as neutrino transport and neutrino-matter
interactions are concerned) three-dimensional (3D) \citep{frywar02}
simulations both confirm that convective overturn indeed occurs in the
$\nu$ heating region and is helpful for shock revival, thus making
explosions possible even when spherically symmetric models fail
\citep{janmue96}. However, despite of strong convective action and a
corresponding enhancement of the efficiency of neutrino energy
transfer to the post-shock matter, even the up to now most realistic
simulations of both non-rotating and rotating progenitor models do not
produce explosions \citep{buras03}.  Multi-dimensional simulations
further demonstrate that vigorous aspherical motion is unavoidable in
the slowly deleptonizing and cooling proto-neutron star
\citep{keil96}.  Both the hot bubble convection \citep{muejan97} and
the convection in the proto-neutron star (see below) as well as the
resulting anisotropic neutrino emission \citep{burrows96, muejan97}
give rise to an interestingly large gravitational wave signal.

Concerning consistent pre-collapse stellar models from evolutionary
calculations of rotating stars, major progress has recently been
achieved by the work of \citet{heger00}, who have performed the up to
now most realistic evolutionary calculations of rotating massive stars
to the onset of core collapse including the effects of rotation,
mixing, transport of angular momentum, most recently, also of magnetic
torques \citep{heger03}. Including the latter the central angular
velocities become of the order of 0.1\,rad/s only.  Despite of these
fairly slow rotation rates the cores are still too fast to lead to
typically expected neutron star natal spins, and angular momentum loss
is considered as a serious problem \citep{woosley03}.  The low iron
core rotation rates, however, suggest that in spite of angular
momentum conservation during core collapse, triaxial
rotation-triggered instabilities need not to be expected in the
collapse simulations. For the time being, we therefore constrain
ourselves to axisymmetric models.

In the following we present the gravitational wave signatures of the
currently most elaborate and detailed core collapse supernova
simulations. One of the considered models starts from an iron core
with a rotation rate at the onset of collapse that is in the ballpark
of predictions from the latest generation of stellar evolution models.
The simulation also contains a state-of-the art description of
neutrino-matter interactions and Boltzmann neutrino transport, a
realistic equation of state, and 2D axisymmetric hydrodynamics taking
into account general relativistic effects. We also present the
gravitational wave signature of a state-of-the-art non-rotating
supernova model. This model, which is based on another, less massive
progenitor, serves the purpose to demonstrate that also from
non-rotating stars interesting gravitational wave signals can be
expected, which are strong enough to be detectable for supernovae that
occur even at large distances in our Galaxy. Finally, we present the
gravitational wave signal expected from the Ledoux convection in a
proto-neutron star at times after the explosion has taken off
($\approx 1\,$s after core bounce).

The paper is organized as follows. In the next section we discuss the
physics employed in our models, and the numerical techniques used to
perform the simulations. In Section\,3 we present the method used to
extract the gravitational wave signature, and in Section\,4 we describe
the results of our investigation.  Finally, in Section\,5 we discuss
the implications of our findings, and give some conclusions.

\section{Input physics and numerical techniques.}
%
For integrating the equations of hydrodynamics we employ the
Newtonian finite-volume code PROMETHEUS \citep{frymue89}.  This
second-order, time-explicit Godunov code is a direct Eulerian
implementation of the Piecewise Pa\-ra\-bol\-ic Method (PPM)
\citep{colwoo84}, and is based on an exact Riemann solver.

We use spherical polar coordinates $(r, \theta, \varphi)$, and assume
axial symmetry.  However, motions in the $ \varphi $-direction are
allowed, and the azimuthal (rotational) velocity may be nonzero.  In
the rotating model the rotation axis coincides with the symmetry axis
at $\theta = 0$. In this model we have also imposed
equatorial symmetry. The 2D computational grid consists of 400
logarithmically spaced zones in radial direction and, if equatorial
symmetry is assumed, of 64 equidistant angular zones covering an
angular range $0 \le \theta \le \pi/2$. If equatorial symmetry is not
assumed, the angular grid consists of 128 equidistant zones in the
angular range $0 \le \theta \le \pi$.

The calculations are performed with the equation of state (EoS) of
\citet{latswe91} using an incompressibility of bulk nuclear matter of
180$\,$MeV. This EoS is based on a compressible liquid-drop model
including nuclei, nucleons, $e^-$, $e^+$, and photons.

Neutrino transport is computed with the Boltzmann solver scheme
described in detail in \citet{ramjan02} and \cite{buras03} for $\nu$
and $\bar\nu$ of all three flavors.  For each angular bin of the
numerical grid the monochromatic moment equations for the radial
transport of $\nu$ number, energy, and momentum are solved. This set
of equations is closed by a variable Eddington factor that is
calculated from the solution of the Boltzmann equation on an angularly
averaged stellar background. Beyond this ``ray-by-ray'' approach we
also take into account the coupling of neighboring rays by lateral
advection terms and $\nu$ pressure gradients \citep{buras03}.  For the
transport an energy grid of 17 geometrically spaced bins is used with
centers from 2$\,$MeV to 333$\,$MeV.

The simulations were done with a Poisson solver that calculates the
gravitational potential from the 2D mass distribution. General
relativistic effects are treated approximately by modifying the
spherical part of the gravitational potential with correction terms
due to pressure and energy of the stellar medium and neutrinos, which
are deduced from a comparison of the Newtonian and relativistic
equations of motion in spherical symmetry \citep{keil96, ramjan02}.
The $\nu$ transport contains gravitational redshift and time dilation,
but ignores the distinction between coordinate radius and proper
radius. Comparison with fully relativistic, one-dimensional
simulations showed that these approximations work well at least when
the deviations of the metric coefficients from unity are moderate
\citep{lieetal03}.

We considered two (solar metallicity) progenitor models with main
sequence masses of $11.2\,M_{\odot}$ and $15\,M_{\odot}$,
respectively \citep{woosley02}. The former progenitor was adopted as
given by the stellar evolution calculation (model s11nr180), while
angular momentum was added to the $15\,M_{\odot}$ progenitor (model
s15r; see also Buras \etal 2003).  The angular frequency of model s15r
was assumed to be constant in the Fe and Si core and to decrease like
$r^{-3/2}$ outside of $1750\,$km ($1.43\,$M$_{\odot}$). This rotation
profile is in agreement with predictions from stellar evolution models
\citep{heger03}. We have adopted a value of $0.5\,$sec$^{-1}$ for the
central angular frequency, which corresponds to an initial value of
$\beta = 0.05\%$, where $\beta$ is the ratio of the rotational energy
and the (absolute) gravitational binding energy.  This is a relatively
small rotation rate compared to those typically assumed in parameter
studies based on rotating polytropes (see, \eg \citet{zwerger97,
dimmelmeier02, ott03}). However, it is of the order of the largest
rotation rates predicted by state-of-the-art evolutionary calculations
of massive stars, and even too fast with respect to observed rotation
rates of young pulsars (\cite{heger03}, see also 
Sect.~\ref{sec:discussion}). Hence, as far as the state of
rotation is concerned, our model is meant to represent not an extreme,
rare case, but is in accord with the conditions in stellar iron cores
as expected from current models for the evolution of rotating massive
stars. Due to the smallness of the assumed rotation rate we regarded
it unnecessary to relax the initial model into rotational equilibrium
(the ratio between the centrifugal force and the gravitational force
is below 1\% in the progenitor model).

We have also analyzed a proto-neutron star model (pns180), which is
the $180^{\circ}$ analogue of the model discussed in \citet{keil96},
which covered an angular region of only $90^{\circ}$ size. Model
pns180 provides complementary information about the gravitational wave
signal expected from core collapse supernovae, namely the wave
emission from the long-lasting ($>1\,$sec) non-radial motion driven by
the deleptonization of the forming neutron star.  The $\sim
1.1\,$M$_{\odot}$ convective core of the proto-neutron star of model
pns180 was formed in the core collapse of a $15\,$M$_{\odot}$ star
\citep{keil96}. Its evolution was followed by \citet{keil97} for a
period of more than 1.2\,sec with a variant of the PROMETHEUS
hydrodynamics code (see above) assuming axial symmetry and including a
flux-limited (equilibrium) neutrino diffusion scheme, which was
applied to each angular bin separately (``1.5D'').  The computational
grid consisted of 100 radial zones extending out to an initial radius
of $58.8\,$km and 120 equidistant angular zones covering an angular
range $0 \le \theta \le \pi$, \ie no equatorial symmetry was assumed.
During the simulated evolution the grid was contracted with the
neutron star shrinking to a final radius of about $20\,$km
\citep{keil96}.

\section{Calculation of the Gravitational Wave Signature} 
%
The (quadrupole) gravitational wave amplitudes and energy spectra
resulting from anisotropic mass motion are computed as described in
\citet{muejan97} (eqs.\,[10]--[12]) using the Einstein quadrupole
formula in the numerically convenient formulation derived by
\cite{blanchet90}, and by standard FFT techniques.  The wave
amplitudes due to the anisotropic neutrino emission are obtained with
the formalism given in \citet{muejan97} (eqs.\,[28]-[31]) which is
based on the work of \citet{epstein78}.

In order to find out whether the gravitational wave signals of our
model stars (assumed to be located at some given distance) are
detectable by terrestrial gravitational wave experiments we have also
computed the corresponding (optimal with respect to the relative
orientation of source and detector) signal-to-noise ratios $S/N$
obtained from matched filtering, which are determined by the relation
\citep{flanagan98}
\begin{equation}
 (S/N)^2 = 4\, \int_{0}^{\infty} \frac{ | \widetilde h ({\nu}) |^2}{
                                        S_n (\nu)} d\nu \, ,
\label{snr}                             
\end{equation}
where $\widetilde h ({\nu})$ is the Fourier transform of the
gravitational wave amplitude $h(t)$ and $S_h (\nu)\,[{\rm Hz}^{-1}]$
is the spectral density of the strain noise in the detector.
Eq.\,(\ref{snr}) can also be written as
\begin{equation}
 (S/N)^2 = 4\, \int_{0}^{\infty} 
               \left[ \frac{ | \widetilde h ({\nu}) | \nu^{1/2} }{
                      h_{\rm rms}(\nu) } \right]^2  d \log \nu \,
\label{snr1}                             
\end{equation}
where $h_{\rm rms}(\nu)\,[{\rm Hz}^{-1/2}] \equiv \sqrt{S_n (\nu)}$ is
the spectral strain sensitivity (see, \eg \citet{fritschel02}), which
gives the rms noise level of the antenna at the frequency $\nu$. The
second expression for the signal-to-noise ratio $S/N$
(Eq.\,\ref{snr1}) is useful when discussing the spectra of the
gravitational wave signals of our models (see Section\,4.2).

\begin{figure}
\plotone{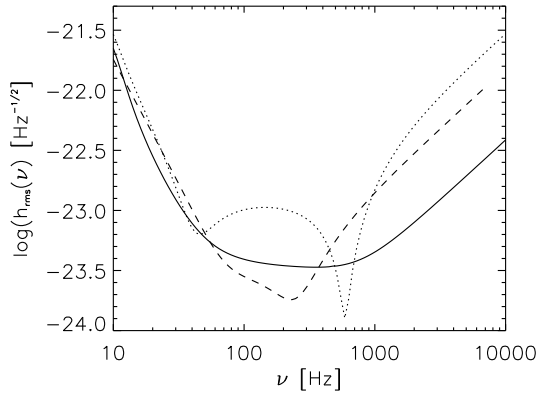}
 \caption{Strain sensitivity curves of the tuned Advanced LIGO
          interferometer (``LIGO\,II'') used in our analysis.
          The solid curve has an increased sensitivity at
          $\sim 1\,$kHz and a rather wide trough of near-maximal
          sensitivity, while the dotted curve has a peak 
          sensitivity of $1.2\,10^{-24}$ at about 700\,Hz, 
          and aligns with the peak in energy in some of our models
          (see Section~\ref{sec:waveresults}).  The third (dashed) 
          sensitivity curve is tuned for sources which emit more
          at low frequencies (the so-called NS-NS-tuned instrument).
 \label{fig:ligo2}} 
\end{figure}

In particular, we have determined the signal-to-noise ratios of our
models both for the LIGO\,I and the tuned Advanced LIGO (``LIGO\,II'')
interferometric gravitational wave detectors \citep{fritschel02}.  For
LIGO\,I we used a fit to the strain sensitivity curve LIGO\,I
SRD\,Goal available from the LIGO web site
www.ligo.caltech.edu/docs/G/G030014-00/.  In case of the tuned
Advanced LIGO interferometer we have computed signal-to-noise ratios
for the three different strain sensitivity curves shown in
Fig.\,\ref{fig:ligo2}.

Concerning the sensitivity curves shown in this plot and
some of our other figures, we point out that both the curves for
LIGO\,II as well as the curve for LIGO\,I are for a {\it single}
interferometer. In the case of LIGO\,I, there are two nominally
identical 4\,km instruments, plus a 2\,km instrument at Hanford. This
increases the net sensitivity by a factor $\sqrt{2}$ in the ideal case
of statistical independence and gaussian noise. For LIGO\,II, there
are 3 interferometers (two at Hanford, one at Livingston). They could
all be tuned the same way for a directed search, or tuned
differently. If all are tuned the same way, the net sensitivity is a
factor $\sqrt{3}$ better at all frequencies than that of a single
interferometer. The signal-to-noise ratios given in this publication
are for the complete detector consisting of two interferometers
(LIGO\,I) or three like-tuned interferometers (LIGO\,II),
respectively.

Assuming axial symmetry and an observer located at an angle $\theta$
with respect to the symmetry axis of the source, the dimensionless
gravitational wave amplitude $h(t)$ is related to the quadrupole wave
amplitude $A^{E2}_{20}$ (measured in units of cm), the lowest-order
non-vanishing term of a multipole expansion of the radiation field
into pure-spin tensor harmonics (see Eq.\,[9] of M\"uller 1997),
according to
\begin{equation}
 h = \frac{1}{8} \sqrt{\frac{15}{\pi}} \sin^2\theta 
     \frac{A^{E2}_{20}}{R} ,
\label{qwa}                             
\end{equation}
where $R$ is the distance to the source. In the following we always
assume $\sin^2\theta = 1$.

\section{Results} 

\subsection{Wave amplitudes}
\begin{figure}
 \plotone{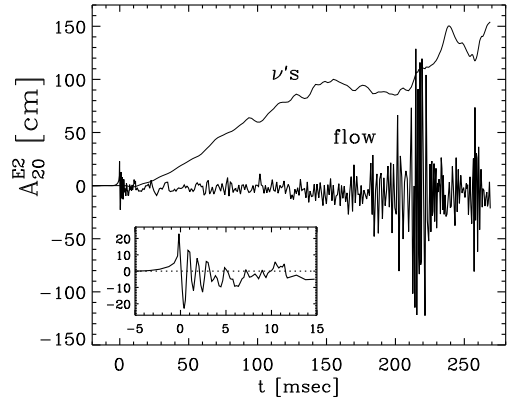}
 \caption{Gravitational wave quadrupole amplitude $A^{E2}_{20}$ vs.\,
          time (post bounce) due to convective mass flow and
          anisotropic neutrino emission (thin line) for the rotating
          delayed explosion model s15r of \citet{buras03}.  The insert
          shows an enlargement of the signal around the time of
          bounce.
 \label{fig:gwa_nurot}} 
\end{figure}

\begin{figure}
 \plotone{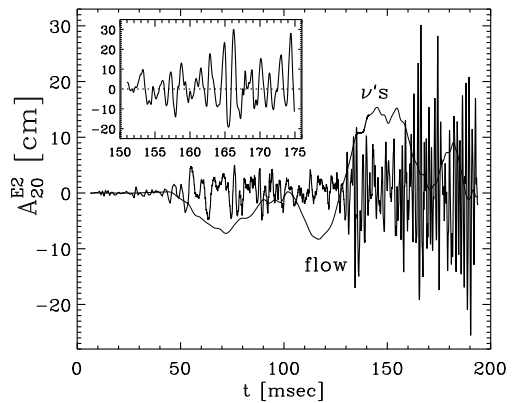}
 \caption{Same as Figure \ref{fig:gwa_nurot} but for the delayed
          explosion model s11nr180 of \cite{buras03}. The insert shows
          an enlargement of the signal between 150\,msec and 175\,msec
          after core bounce.
 \label{fig:gwa_nu180}}
\end{figure}

\begin{figure}
 \plotone{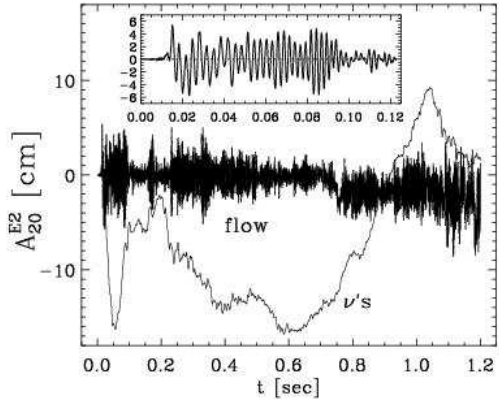}
  \caption{Same as Fig.\,\ref{fig:gwa_nurot} but for the proto-neutron
           star model pns180 of \citet{keil97}.  Note that no bounce
           occurs in this model, which follows the neutrino cooling
           phase of the nascent neutron star. Time is measured from
           the start of the simulation at about 20\,msec after stellar
           core collapse.
 \label{fig:gwa_BF2D}}
\end{figure}

The quadrupole wave amplitudes of models s15r, s11nr180 and pns180 are
shown in Figs.\,\ref{fig:gwa_nurot}, \ref{fig:gwa_nu180} and
\ref{fig:gwa_BF2D}, respectively. Besides the waveform resulting from
the convective flow, each figure shows the signal due to the
anisotropic neutrino emission. Inserts further illustrate details of
the wave signal for chosen periods of time, and demonstrate that the
wave signal consists of many quasi-periodic variations on time scales of
a few milliseconds which remain unresolved when the whole time
evolution of the models is shown.

\begin{figure}
 \plotone{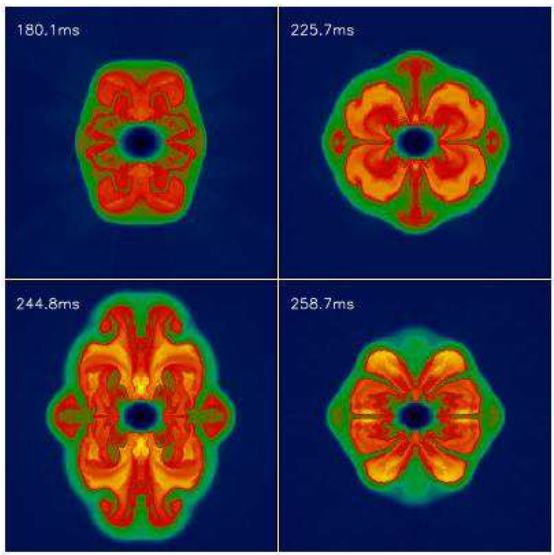} 
 \caption{Four snapshots of the entropy distribution of the rotating
          delayed explosion model s15r of \citet{buras03}. The side length
          of the plots is $600\,$km, and the numbers in the top
          left corners give post-bounce times. Bright regions are high
          entropy bubbles, the shock is visible as the sharp, deformed
          discontinuity, and the flattened proto-neutron star is the
          low-entropy oblate ellipse at the center.
 \label{fig:rampp-rotmod}}
\end{figure}

The quadrupole wave amplitude $A^{E2}_{20}$ (Eq.\,\ref{qwa}) of model
s15r at bounce is of the order of 20\,cm showing (see insert of
Fig.\,\ref{fig:gwa_nurot}) the typical prominent spike with subsequent
ring-down of a type\,I gravitational wave signal of a rotating core,
which bounces due to the stiffening of the equation of state around
nuclear matter density \citep{moenchmeyer91}. The post-bounce wave
amplitude stays below $\sim 10\,$cm until about 150\,msec when
convective mass motions driven by neutrino heating in combination with
the increasing rotation rate of the post-shock region amplify the wave
signal. The growing rate of rotation of the post-shock region is
caused by the accretion of matter with higher angular momentum (as the
initial rotation profile is constant to the edge of the Fe core)
through the shock, and the specific angular momentum increasing even
beyond the edge of the iron core.  The post-shock region starts to
exhibit large-scale non-radial pulsations
(Fig.\,\ref{fig:rampp-rotmod}) which produce a long-time ($\sim10 -
20\,$msec) variability on top of the rapidly varying ($\sim\,$msec)
wave signal of up to 130\,cm amplitude.

\begin{figure}
   \plotone{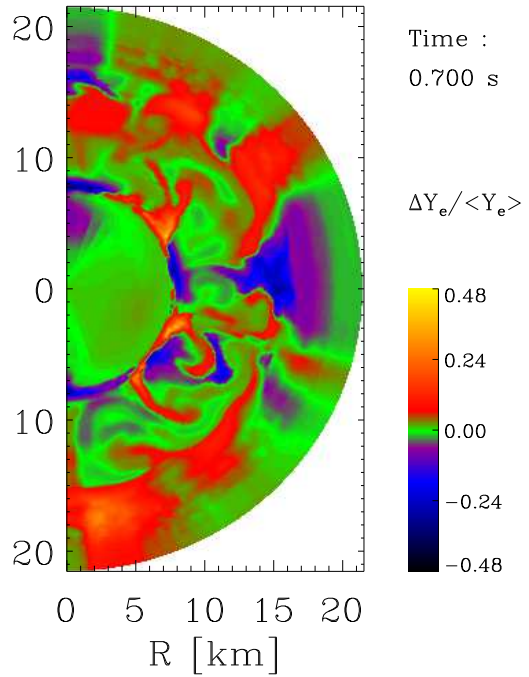} 
   \caption{Relative variation of the proton-to-baryon ratio (electron
            number fraction) due to convective mass flow in the
            proto-neutron star model pns180 of \citet{keil97} 0.7\,sec
            after core bounce. Proton-rich matter rises, while
            neutron-rich matter sinks inward.
\label{fig:BF2D_yevar}}
\end{figure}

The non-rotating, and hence initially spherically symmetric model
s11nr180 shows, as expected, no signal at bounce
(Fig.\,\ref{fig:gwa_nu180}). Until 50\,msec after bounce the wave
amplitude of this model remains tiny ($<1\,$cm). Only when the
neutrino-driven convection sets in at around $40\,$msec to $50\,$msec,
the wave amplitude grows to the size of a few centimeters. Eventually,
as the convective activity becomes stronger, the maximum (absolute)
amplitudes rise to values of several 10 centimeters.

Both models, the rotating (s15r) as well as the non-rotating one
(s11nr180), do not explode within the displayed time intervals.
After an explosion has started we expect a slow decay
of the convective activity behind the expanding supernova
shock, and hence a declining gravitational wave amplitude from this
region over a period of a few $100\,$msec. The gravitational wave signal
afterwards will be dominated by the emission from convection {\em
inside} the cooling nascent neutron star.

The proto-neutron star model exhibits a gravitational wave signal
consisting of rapid ($\sim\,$msec) quasi-periodic temporal variations of
roughly constant amplitude ($\sim \,$cm) modulated on time scales of
$\sim 100\,$ms (Fig.\,\ref{fig:gwa_BF2D}). While the former variations
reflect mass motions, and non-radial proto-neutron star oscillations
occurring on the hydrodynamic (sound crossing) time scale, the
long-time signal modulations are caused by the varying strength of the
activity due to Ledoux convection in the proto-neutron star
(Fig.\,\ref{fig:BF2D_yevar}).

The wave amplitude associated with anisotropic neutrino emission shows
much less time structure and slower temporal variation than that
produced by non-radial mass motion. It is characterized by an overall
rise of the wave amplitude in case of the rotating model s15r, a
quasi-periodic oscillation with a time scale of $\sim 50\,$msec in case
of the non-rotating model s11nr180, and an even slower temporal
variation ($0.2$---$0.4\,$sec) in case of the proto-neutron star
model pns180.

In case of the rotating model (s15r) the temporal variation of the
quadrupole wave amplitude of the neutrinos reflects the time scale of
the large-scale, non-radial pulsations exhibited by the post-shock
region ($\sim 10$ to $\sim 20\,$msec; Fig.\,\ref{fig:rampp-rotmod}),
while the short-time variations ($\sim\,$\,msec) of the wave amplitude
of the mass flow are caused by the convective activity in the hot
bubble region.  The overall rise of the wave amplitude of the
neutrinos is a consequence of the steadily increasing rotation rate
(accretion of matter with higher angular momentum) and thus growing
rotational flattening of the nascent neutron star. In the
non-rotating model (Fig.\,\ref{fig:gwa_nu180}) and the proto-neutron
star model (Fig.\,\ref{fig:gwa_BF2D}) the temporal variations of the
wave amplitudes of the neutrinos are caused by large-scale, localized
neutrino-emitting downflows through the neutrino-heated bubble, and by
global changes in the location and shape of the convective emission
regions, respectively.

\subsection{Wave spectra and signal-to-noise ratios}
\label{sec:waveresults}
\begin{figure}
 \plotone{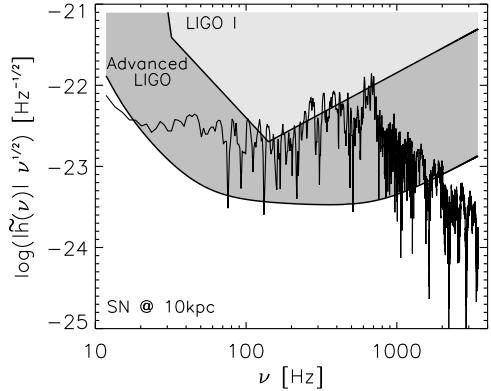}
 \caption{Spectral energy distribution of the quadrupole radiation due
          to convective mass flow and anisotropic neutrino emission
          for the rotating delayed explosion model s15r of
          \citet{buras03} as a function of frequency of the emitted
          gravitational radiation for a supernova at a distance of
          10\,kpc.
 \label{fig:gwa_fft_nurot_sum}}
\end{figure}

\begin{figure}
 \plotone{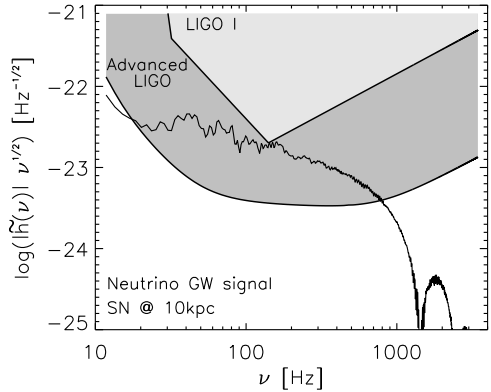}
 \caption{Same as Fig.\,\ref{fig:gwa_fft_nurot_sum} but showing only
          the contribution due to anisotropic neutrino emission.
 \label{fig:gwa_fft_nurot_nue}}
\end{figure}

\begin{figure}
 \plotone{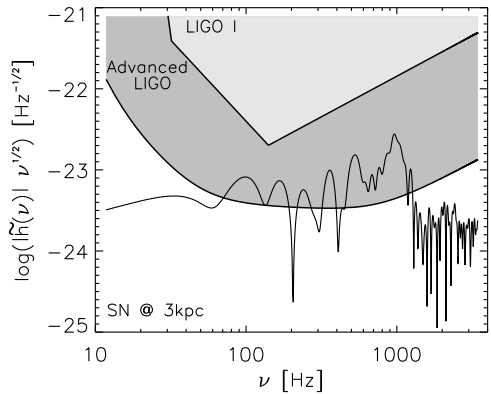}
 \caption{Same as Fig.\,\ref{fig:gwa_fft_nurot_sum} but for the bounce
          signal only, and a supernova at a distance of 3\,kpc.
 \label{fig:gwa_fft_nurot_bnc}}
\end{figure}

\begin{figure}
 \plotone{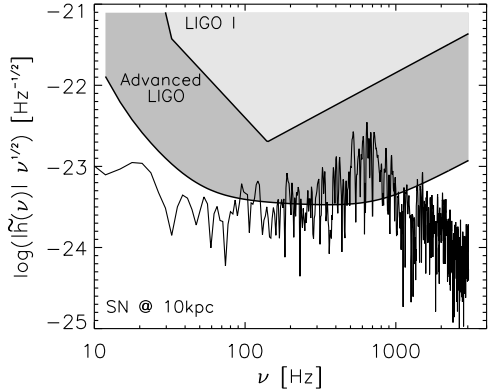}
 \caption{Same as Figure \ref{fig:gwa_fft_nurot_sum} but for the
          delayed explosion model s11nr180 of \cite{buras03}.
 \label{fig:gwa_fft_nu180_sum}}
\end{figure}

\begin{figure}
 \plotone{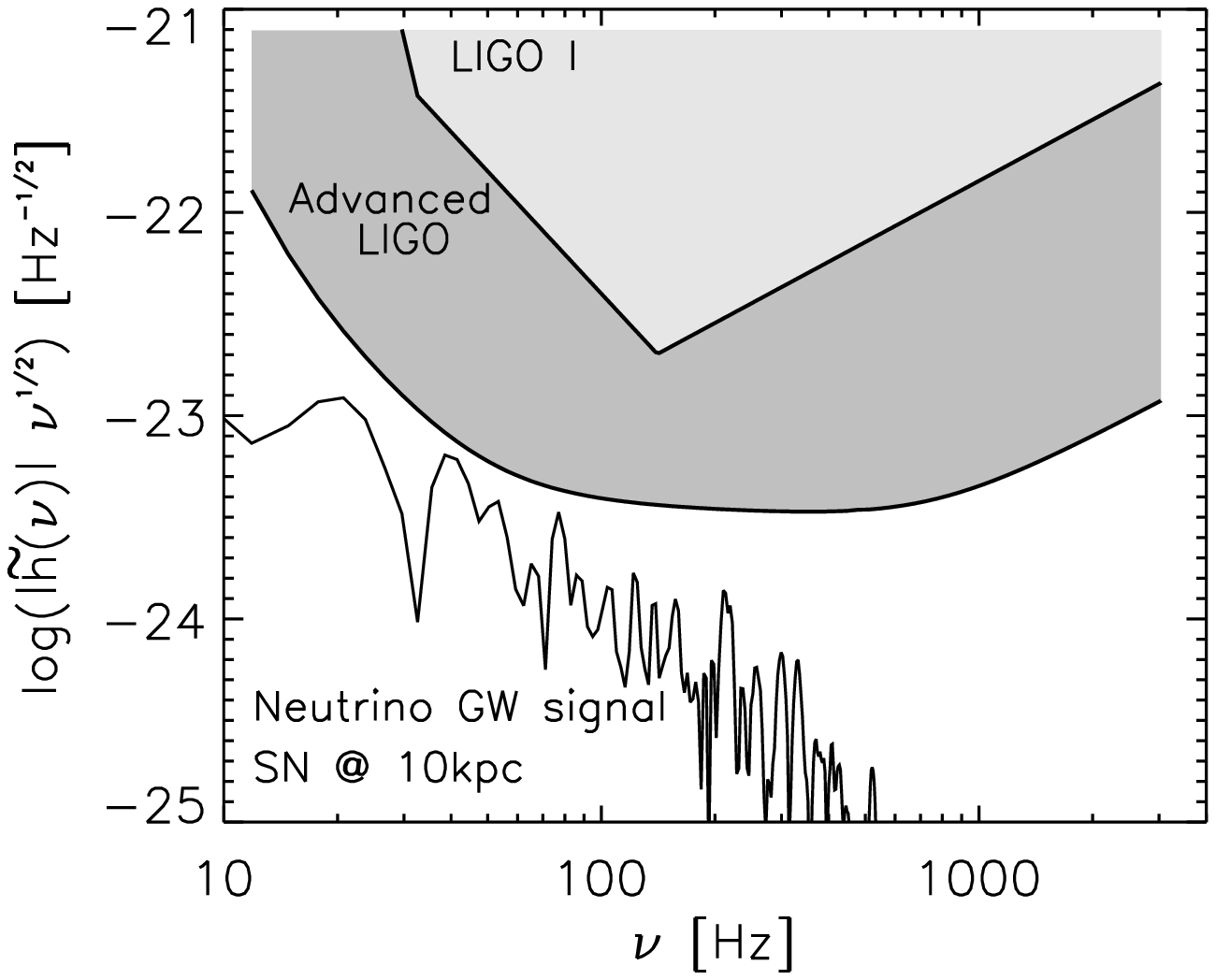}
 \caption{Same as Figure \ref{fig:gwa_fft_nurot_nue} but for the
          delayed explosion model s11nr180 of \cite{buras03}.
 \label{fig:gwa_fft_nu180_nue}}
\end{figure}

\begin{figure}
 \plotone{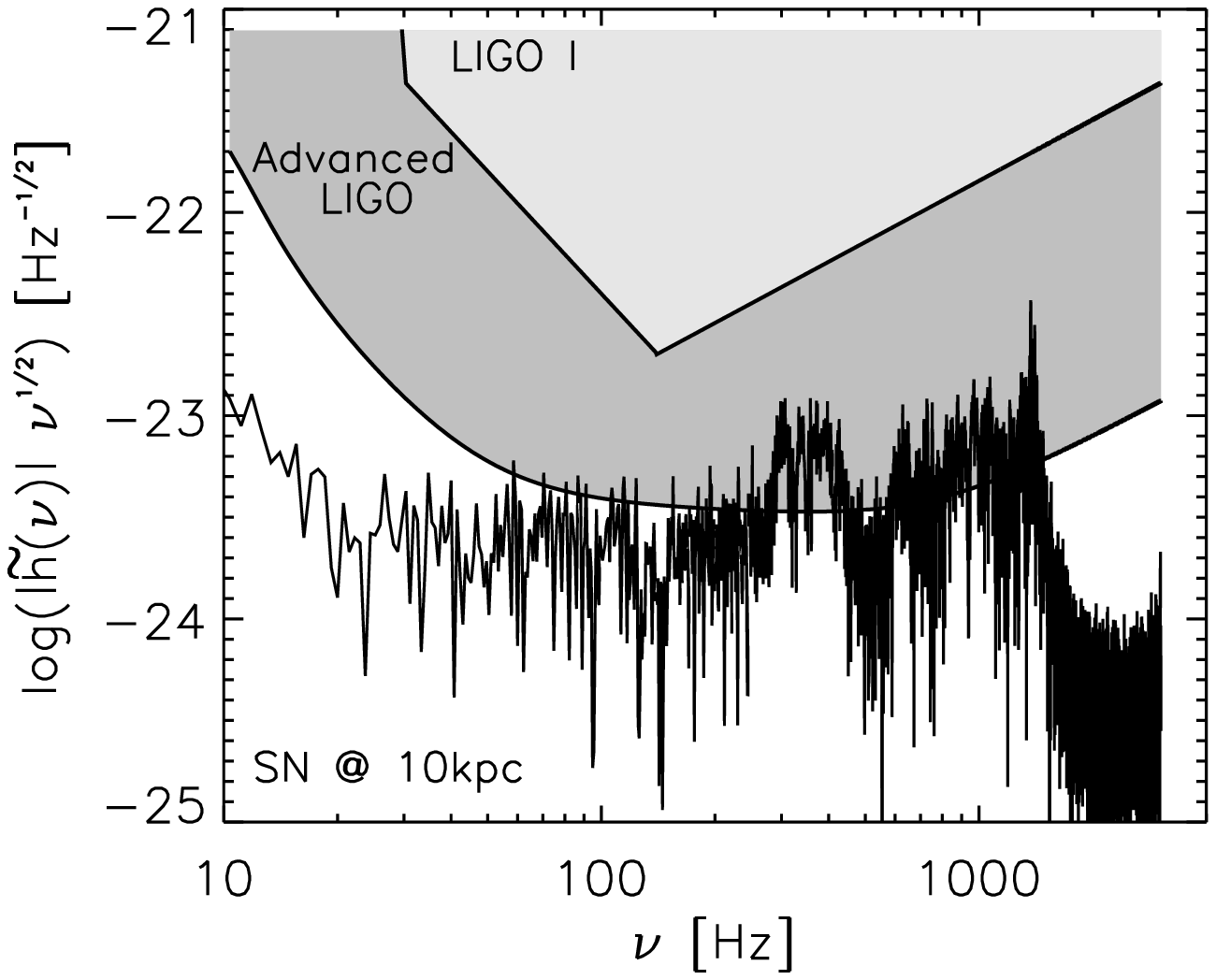}
 \caption{Same as Fig.\,\ref{fig:gwa_fft_nurot_sum} but for the
          proto-neutron star model pns180 of \citet{keil97}.
 \label{fig:gwa_fft_BF2D_sum}}
\end{figure}

\begin{figure}
 \plotone{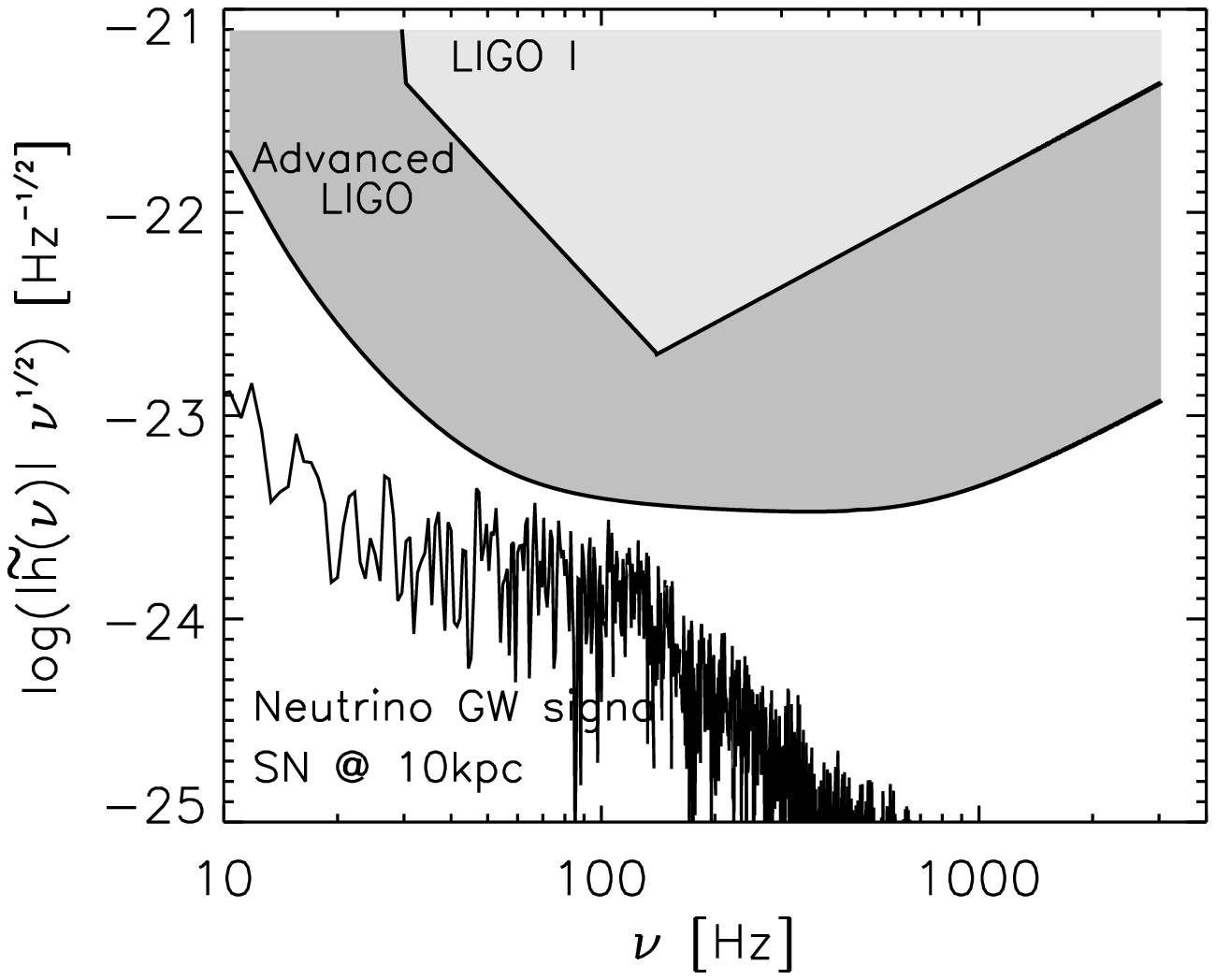}
 \caption{Same as Fig.\,\ref{fig:gwa_fft_nurot_nue} but for the
          proto-neutron star model pns180 of \citet{keil97}.
 \label{fig:gwa_fft_BF2D_nue}}
\end{figure}

In Figs.\,\ref{fig:gwa_fft_nurot_sum} to \ref{fig:gwa_fft_BF2D_nue} we
show the (logarithm) of the quantity $| \widetilde h (\nu) |
\nu^{1/2}$ appearing in the numerator of Eq.\,(\ref{snr1}). Each
figure further gives $h_{\rm rms} (\nu)$, the rms noise level (see
denominator of Eq.\,\ref{snr1}) of a LIGO\,I and LIGO\,II single
interferometer (cf.~Sect.~3) as a function of the frequency $\nu$,
the LIGO\,II curve being the solid curve of Fig.\,\ref{fig:ligo2}.
Both quantities determine the signal-to-noise ratio ($S/N$), which is
given for all models and both antennae in Table\,\ref{tbl_snr} using
again the solid sensitivity curve of Fig.\,\ref{fig:ligo2}. We remind the
reader that the $S/N$ values are computed for the complete LIGO\,I and
LIGO\,II detectors as explained in Sect.~3.


The spectrum of the complete wave train of model s15r caused by both
anisotropic mass flow and neutrino emission is displayed in
Fig.\,\ref{fig:gwa_fft_nurot_sum} for a supernova at a distance of
$10\,$kpc. It rises to a broad peak between $\sim 600\,$Hz and $\sim
700\,$Hz and then declines rapidly at higher frequencies.  The figure
indicates that while the signal-to-noise ratio is probably too small
for this event to be detectable by LIGO\,I, it could be very well
detected by LIGO\,II. This is confirmed by the signal-to-noise ratios
computed from Eq.\,(\ref{snr1}) (using the solid sensitivity
curve of Fig.\,\ref{fig:ligo2}), which are 3.7 and 67 for the LIGO\,I
and LIGO\,II detectors, respectively (Table\,\ref{tbl_snr}).  Even at
a distance of $100\,$kpc the signal-to-noise ratio is still $\approx
7$ for LIGO\,II, \ie it is probably sufficiently large for a detection
of the event \citep{flanagan98}.  Comparing
Fig.\,\ref{fig:gwa_fft_nurot_sum} with
Fig.\,\ref{fig:gwa_fft_nurot_nue}, which shows the spectrum resulting
from anisotropic neutrino emission alone, one recognizes that the
low-frequency part of the spectrum (below $\sim 100\,$Hz) is dominated
by the contribution of the neutrinos, because their quadrupole wave
amplitude varies on much longer time scales than that of the mass flow
(Fig.\,\ref{fig:gwa_nurot}).
 
When we Fourier transform just the bounce signal (insert in
Fig.\,\ref{fig:gwa_nurot}), \ie the wave signal up to
$15\,$msec past core bounce, we find that its spectrum peaks at a
somewhat higher frequency of $\sim 900\,$Hz
(Fig.\,\ref{fig:gwa_fft_nurot_bnc}) than that of the complete
signal. The signal-to-noise ratios drop to 0.7 (LIGO\,I) and 14
(LIGO\,II) even when the source is located at a smaller distance of
$3\,$kpc (Table\,\ref{tbl_snr}).

The spectrum of the non-rotating model s11nr180
(Fig.\,\ref{fig:gwa_fft_nu180_sum}) is qualitatively quite similar to
that of the rotating model s15r (Fig.\,\ref{fig:gwa_fft_nurot_sum}),
but the corresponding signal-to-noise ratios are much smaller for a
source located at the same distance ($10\,$kpc): $S/N=0.6$ (LIGO\,I)
and $S/N=13$ (LIGO\,II), respectively (Table\,\ref{tbl_snr}). Thus,
while a non-rotating supernova (model s11nr180) is detectable with
LIGO\,II up to a distance of $\sim 15\,$kpc, a massive star rotating
at a rate in the ballpark of predictions by state-of-the-art stellar
evolution models will produce a gravitational wave supernova signal
which can clearly be seen by LIGO\,II throughout the Galaxy and
in the Magellanic Clouds, and by LIGO\,I if the supernova occurs less
than $\sim 5\,$kpc away from Earth, a distance limit that encompasses
all historical Galactic supernovae.  As in case of the rotating model,
the low frequency part of the spectrum of model s11nr180 is dominated
by the contribution from anisotropic neutrino emission
(Fig.\,\ref{fig:gwa_fft_nu180_nue}) because of the slower temporal
variation of the corresponding quadrupole wave amplitude
(Fig.\,\ref{fig:gwa_nu180}).



\begin{deluxetable}{l|r|rr|r}
\tablecolumns{5}
\tablewidth{0.8\columwidth}
\tablecaption{ Signal-to-noise ratios\tablenotemark{a} for LIGO\,I and II.
\label{tbl_snr}}
\tablehead{ \colhead{model}  & \colhead{D\,[kpc]\tablenotemark{b}} & 
            \colhead{S/N\,I} & \colhead{S/N\,II} &
            \colhead{$E_{\rm GW}\,[M_\odot c^2]$\tablenotemark{c}} }
\startdata
s15r                       & 10 & 3.7 & 67  &  $3.0\,10^{-09}$ \\
s15r bnc\tablenotemark{d}  &  3 & 0.7 & 14  &  $3.6\,10^{-11}$ \\
s11nr180                   & 10 & 0.6 & 13  &  $1.9\,10^{-10}$ \\
pns180                     & 10 & 0.4 &  8  &  $1.6\,10^{-10}$ \\[-5pt]
\enddata
\tablenotetext{a}{for complete instruments with multiple antennae}
\tablenotetext{b}{source distance}
\tablenotetext{c}{radiated gravitational wave energy}
\tablenotetext{d}{only bounce signal of model s15r}
\end{deluxetable}

According to our results the ``long time'' ($\ga 1\,$sec) gravitational
wave signal produced by all core collapse supernova events due to
convective activity inside the proto-neutron star is detectable with
LIGO\,II ($S/N=8$) up to a distance of $10\,$kpc
(Fig.\,\ref{fig:gwa_fft_BF2D_sum} and Table\,\ref{tbl_snr}). The
spectrum displays a narrow dominant peak at $\sim 1.3\,$kHz, and two
additional, weaker broad maxima centered at $1\,$kHz and $\sim
350\,$Hz. Anisotropic neutrino emission again dominates the spectrum
at frequencies below $\sim 100\,$Hz
(Fig.\,\ref{fig:gwa_fft_BF2D_nue}). We point out here that the
convective activity in the proto-neutron star may continue beyond the
time period covered by our simulation, perhaps lasting more than
$10\,$sec until the deleptonization is complete and the convective
motion gets damped by viscous effects. Assuming that the amplitude and
the spectrum of the gravitational wave emission does not change
considerably during this time, the signal-to-noise ratios quoted for
this model in Table\,\ref{tbl_snr} may increase by a factor of $\ga
10$, as they depend linearly on the size of the Fourier amplitude
which grows proportional to the length of the signal.

The signal-to-noise ratios given in Table\,\ref{tbl_snr} for the
complete tuned Advanced LIGO depend somewhat, but not extremely, on
the choice of the sensitivity curve. In case of model s15r the
signal-to-noise ratio increases only slightly from 67 to 71 and 70
when using the dotted and dashed curves of Fig.\,\ref{fig:ligo2},
respectively. The corresponding signal-to-noise ratios for model
s11nr180 are 13, 17 and 10, and those for model pns180 are 8, 5 and
8. Hence, for the latter two models tuning of the senstivity curve for
source optimization can increase the S/N values by up to 70\%.

\subsection{Amount of Radiated Energy}
The energy radiated in the form of gravitational waves is given by
(see \eg Eqs.\,[22], [26] and [28] in \citet{zwerger97})
\begin{eqnarray}
 E_{\rm GW} &=& \frac{c^3}{G} \frac{1}{32\pi}
                \int_{-\infty}^{\infty} 
                \left( \frac{d A^{E2}_{20}(t)}{dt} \right)^2 dt
\\
            &=& \frac{c^3}{G} \frac{1}{16\pi}
                \int_{0}^{\infty} | \widetilde A^{E2}_{20}(\nu) |^2 
                     \nu^2 d\nu \, ,
\label{egw}                             
\end{eqnarray}
where $\widetilde A^{E2}_{20}$ is the Fourier transform of
$A^{E2}_{20}$, and $c$ and $G$ are the speed of light in vacuum and
Newton's gravitational constant, respectively. We used both
expressions above to compute the radiated energy, and found for all
three models very good agreement with numerical differences only in
the few percent range.

Our analysis implies that non-spherical core collapse supernovae
rotating with rates in the ballpark of predictions by state-of-the-art
evolutionary calculations of massive stars emit an energy of a few
$10^{-9} M_\odot c^2 $ ($ M_\odot$ is the solar mass) in the form of
gravitational radiation (Table\,\ref{tbl_snr}). This number might
increase for stars more massive than the considered $15\,M_\odot$
progenitor, and is reduced by roughly one order of magnitude when
non-rotating cores or the emission of convective proto-neutron stars
are considered. Anisotropic neutrino emission does not contribute much
to the amount of radiated energy, as the corresponding quadrupole wave
amplitudes show a slower and smaller temporal variation than those due
to the non-spherical mass flow.

\section{Discussion and conclusions}
\label{sec:discussion}
We have presented the gravitational wave signature of core collapse
supernovae using the presently most realistic input physics
available. Our analysis is based on state-of-the-art progenitor models
of rotating and non-rotating massive stars, whose axisymmetric core
collapse is simulated by integrating the hydrodynamic equations
together with the Boltzmann equation for the neutrino transport
including an elaborate description of neutrino interactions, and a
realistic equation of state.  We have computed the quadrupole wave
amplitudes, the Fourier wave spectra, the amount of energy radiated in
form of gravitational waves, and the signal-to-noise ratios for the
LIGO\,I and LIGO\,II detectors resulting both from non-radial mass
motion and anisotropic neutrino emission.

The results show that the dominant contribution to the gravitational
wave signal is not produced by the core bounce itself, but instead by
the neutrino-driven convection in the post-shock region lasting for up
to several hundred milliseconds post bounce. This finding is
qualitatively different from that of previous core collapse
simulations which have been performed with greatly simplified
parameterized models involving (i) a polytropic equation of state,
and/or (ii) a simplified description of weak interactions and neutrino
transport or none at all, and/or (iii) parameterized pre-collapse
stellar models, and/or (iv) Newtonian gravity. We further find that
the contribution to the signal from anisotropic neutrino emission is
dominant at frequencies below $\sim 100\,$Hz, while non-spherical mass
flow dominantly contributes in the range between $\sim 300\,$Hz and
$\sim 1200\,$Hz. Besides the neutrino-driven convection in the
post-shock region, the Ledoux convection inside the deleptonizing
proto-neutron star also gives rise to an important gravitational wave
signal from core collapse supernovae. 

For a stellar iron core rotating with a central angular frequency of
$0.5\,$s$^{-1}$, which is relatively small compared to the rotation
rates typically assumed in parameter studies based on rotating
polytropes (see, \eg \citet{zwerger97, dimmelmeier02, ott03}), but
which is of the order of the largest rotation rates predicted by
state-of-the-art evolutionary calculations of massive stars
\citep{heger03}, the core bounce signal would be detectable ($S/N
\ga 7$) with LIGO\,II for a supernova up to a distance of $\sim
5\,$kpc.  However, the signal caused by neutrino-driven convection is
observable with LIGO\,II up to a distance of $\sim 100\,$kpc, and even
with LIGO\,I to a distance of $\sim 5\,$kpc.

Due to the smallness of the assumed initial rotation rate we do not
expect any significant growth of non-axisymmetric instabilities (\eg
bar modes) even after core collapse when angular momentum conservation
leads to a spin-up of the core. For the same reason there also occurs
no centrifugal hang-up, but a regular bounce at about nuclear matter
density. Consequently, the gravitational wave (bounce) signal is not
of type\,II consisting of several distinct spikes, but of type\,I with
one prominent spike at bounce followed by a ring-down
\citep{moenchmeyer91}.

As far as the initial state of rotation is concerned, our model
represents not an extreme, rare case, but is in accord with more
typical iron core conditions as expected from the evolution of
rotating massive stars. We point out here that most models used in
previous investigations have too large rotational rates in comparison
to those observed for newborn pulsars (see \eg \citet{kotake03}). Even
our initially quite slowly rotating model will produce a pulsar
spinning with a period, $P_{\rm rot}$, between 1 and $2\,$msec
(depending on its final radius of 10--$15\,$km and assuming
angular momentum conservation in the neutron star after the end of 
our simulations), which is not far from the breakup limit.
Millisecond pulsars have rotational energies $\sim
5\,10^{52}\,{\rm erg}\, (P_{\rm rot}/1\,{\rm msec})^{-2} (R/10\,{\rm
km})^2 M/(1.5\,M_{\odot})$. If such rapid rotators have to be slowed
down to match the rotation rates of observed young pulsars, which is
difficult to achieve on short time scales \citep{woosley03}, the
question arises where all this rotational energy has gone, if it is
not observable in kinetic energy or electromagnetic radiation from the
supernova or pulsar?  R-modes do also no longer appear as a likely
energy drain \citep{arras03}.

If the core is non-rotating (model s11nr180) its gravitational wave
signal due to aspherical motion in the neutrino-heated bubble and due
to anisotropic neutrino emission is detectable ($S/N \ga 8$) with
LIGO\,II up to a distance of $\sim 15\,$kpc, while the signal from the
Ledoux convection (model pns180) in the deleptonizing proto-neutron
star can be measured at least to a distance of $\sim 10\,$kpc. Both
kinds of signals are generically produced in any core collapse
supernova.

As our models are axisymmetric, the results may change when 3D effects
associated with the convection inside the proto-neutron star and in
the hot bubble region are included. Both phenomena are genuinely three
dimensional. In this respect the simulations of \citet{frywar02}, and
their analysis of the gravitational wave emission \citep{fryer03} may
provide an answer to this uncertainty in our results. These authors
find that the size and dynamics of the convective structures are
similar to those in axisymmetric simulations.
 
Finally, we want to mention another mechanism which causes a
gravitational wave signal in core collapse explosions.  Recent
simulations by \citet{scheck03} indicate that the proto-neutron star
is accelerated during the explosion at a rate of up to $\sim
3000\,$km/sec$^2$, resulting in large kick velocities in some models.
The acceleration can be caused either by asymmetric accretion onto the
proto-neutron star, or by the pull of the gravitational potential of
matter distributed asymmetrically in the surrounding hot bubble
region. Due to the erratic behavior of the accretion downflows the
acceleration can temporarily change sign leading to an additional
oscillatory component superimposed on or preceding the steady
acceleration component.

An order of magnitude estimate of the resulting scale of the
gravitational wave signal can be obtained by approximating the
proto-neutron star by a point mass $M$ (initially at rest), which \eg
moves in positive $z$-direction starting at $z=0$.  If the point mass
experiences a constant acceleration $b$ for a time $t$, its
displacement is given by $z(t) = b/2\, t^2$.  The only non-vanishing
(Cartesian) component of the time-dependent quadrupole moment
(measured with respect to the origin of the coordinate system) of such
a moving point mass is $Q_{zz} = 2/3\, M z^2(t) = 1/6\, M b^2 t^4 $. The
lowest (quadrupole) order contribution to the gravitational wave
luminosity $L \equiv d E_{\rm GW} / dt$ of such a moving point mass is
due to the temporal variation of its quadrupole moment
\citep{thorne80}, and is given by $L_{\rm kick} = G/(5c^5)\,
\langle (d^3 Q_{zz} / dt^3)^2 \rangle = 16G/(5 c^5)\, M^2 b^4 t^2$,
which can be written as
\begin{equation}
 L_{\rm kick} = 2.0\,10^{-19}\, \left(\frac{M}{M_{\odot}}\right)^2\,  
                b_2^4\, t^2 \, \left[\frac{M_{\odot}c^2}{{\rm sec}}\right] \, ,
\label{eq:gwkick}
\end{equation}
where $b_2$ is the acceleration in units of $10^2\,$km/sec$^2$, and
$t$ the time of acceleration, measured in seconds.  If the motion of
the point mass (initially at rest at $z=0$) is caused by a periodic
acceleration of amplitude $A$ and time scale $\tau$, \ie $d^2 z(t) /
dt^2 = A \cos (2\pi t / \tau)$, the resulting gravitational wave
luminosity is given by $L_{\rm osc} = 2G/(9\pi^2c^5)\, M^2\, A^4\,
\tau^2$, or by
\begin{equation}
 L_{\rm osc} = 1.4\,10^{-19}\, \left(\frac{M}{M_{\odot}}\right)^2\, A_3^4\,
               \tau_{-1}^2\, \left[\frac{M_{\odot}c^2}{{\rm sec}}\right] \, ,
\label{eq:gwosc}
\end{equation}
where $A_3$ is the acceleration in units of $10^3\,$km/sec$^2$, and
$\tau_{-1}$ the oscillation time scale, measured in units of
$10^{-1}\,$sec. From their simulations \citet{scheck03} infer $b_2 \la
5$, $t \approx 1$, $A_3 \la 3$, and $\tau_{-1} \approx 1$ for the
proto-neutron star of mass $M \sim M_{\odot}$. Hence, the
gravitational wave signal from the acceleration giving rise to the
kick of the proto-neutron star is quite small ($L_{\rm kick} \la
10^{-16}\,[M_{\odot}c^2/{\rm sec}]$), and the signal from the
oscillatory motion of the proto-neutron star is even smaller ($L_{\rm
osc} \la 10^{-17}\,[M_{\odot}c^2/{\rm sec}]$). The former signal is
expected to have a frequency of typically $\sim\,$Hz, and the latter
one should be dominant at about 10 to $20\,$Hz.

Both signals are very small compared to the signal caused by the
bounce of a rotating core, because the accelerations encountered
during core bounce are much larger. The accelerations, which cause the
varying quadrupolar deformation of the rotating core, are of the order
$10^6\,$km/sec$^2$, \ie they are roughly a factor of $10^3$ larger
than those producing the kick or the oscillatory motion of the
proto-neutron star. These accelerations act for only $\sim\,10\,$msec
(the damping time scale of the post-bounce ringing; see \eg the insert
in Fig.\,\ref{fig:gwa_nurot}), which is a factor $\sim\,100$
($\sim\,10$) shorter than $t$ ($\tau_{-1}$), and involve a mass of only
$\sim 0.1\,M_{\odot}$, which is a factor $\sim\,10$ smaller than $M$.
Nevertheless, as $L$ scales bi-quadratically with the acceleration and
only quadratically with the mass and time scale (see
Eqs.\,\ref{eq:gwkick} and \ref{eq:gwosc}), the gravitational wave
luminosity of rotational core bounce exceeds that of the proto-neutron
star (kick and oscillatory) motion by a factor $\sim\,10^6$ to
$\sim\,10^8$.

\acknowledgments{
The authors would like to thank W.\,Keil for computing the
proto-neutron star model used in our analysis.  Support by the
Sonderforschungsbereich 375 on ``Astroparticle Physics'', and the
Sonderforschungsbereich/Transregio 6020 on ``Gravitational Wave
Astronomy'' of the Deutsche Forschungsgemeinschaft is acknowledged.
The simulations were performed at the Rechenzentrum Garching (RZG) of
the Max-Planck-Society.
}

\end{document}